# Sharp enhancement on thermoelectric figure-of-merit of post-transition metal chalcogenides (PTMCs) using heterostructures with Mexican-hat valence band

Marcel S. Claro[*,1,2]

[1] INL-International Iberian Nanotechnology Laboratory, Av. Mestre José Veiga s/n, 4715-330 Braga, Portugal.

[2] Centro Singular de Investigación en Química Bioloxica e Materiais Moleculares (CiQUS), Departamento de Química-Física, Universidade de Santiago de Compostela, Santiago de Compostela 15782, Spain

e-mail: marcel.santos@usc.es

**ABSTRACT:** Post-transition metal chalcogenides (PTMCs) such as GaSe, GaS, InSe, and InS have been proposed as promising thermoelectric materials due to low lattice conductivity, originating from the atomically layered structure, high Seebeck coefficient, and the anticipation that its figure-of-merit be improved when thinned to few-layers as the band structure turns into Mexican-hat valence band (MHVB). Here we show by *ab initio* calculations that the MHVB should be present even on thick films of InSe/GaSe type-II heterostructures, and a 50% enhancement on thermoelectric figure-of-merit zT at room-temperature is expected when compared with bulk InSe.

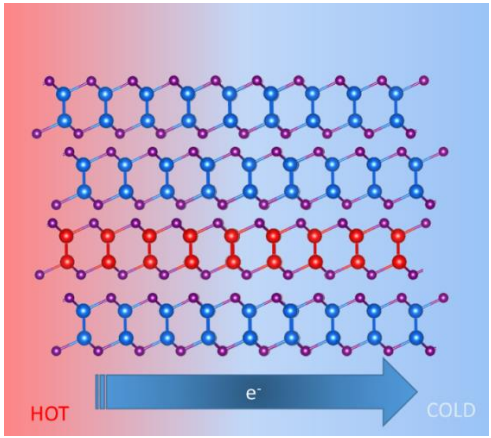

Thermoelectric (TE) materials and devices can directly convert heat to electricity based on the Seebeck effect, and are key enabling technologies of new clean (carbon-free) sources of energy and to harvest energy which is usually wasted on inefficient equipment and energy generation.[1–3] Nonetheless, The main obstacle to the wide adoption of TE devices is their low efficiency. The efficiency of thermoelectric materials is measured by the figure-of-merit $zT(T) = S^2\sigma T/(\kappa_e + \kappa_L)$, where S is the Seebeck coefficient, σ is the electrical conductivity, T is the absolute temperature, $\kappa_L$ and $\kappa_e$ are the lattice and electronic thermal conductivities, respectively. Increase zT value has been proved challenging over the years, as it requires simultaneously a high Seebeck coefficient and electrical conductivity but low thermal conductivity, when high electrical conductivity actually has a detrimental effect on the electron thermal conductivity (Wiedemann-Franz law). Various methods are proposed to enhance the TE efficiency, including band engineering to improve the power factor ($S^2\sigma$), for example, using low dimensional semiconductors to obtain step-like and delta-like density of states (DOS).[4] Another typical example is use materials with Mexican-hat valence band (MHVB), present in the most efficient TE materials, e.g., $Bi_2Se_3$ and $Bi_2Te_3$.[5,6] This band profile combines the large DOS and large transport velocity due to the flatness of the band. Therefore, enhancing at the same time all the TE properties.

Mexican-hat dispersions are relatively common in few-layer two-dimensional materials. *Ab initio* studies and ARPES direct measurements have found Mexican-hat dispersions in the valence band of many few-layer Post-transition metal chalcogenides (PTMCs) materials such as GaSe, GaS, InSe, and InS.[7–9] These materials are interesting for electronic and photonic application and also have been proposed as promising TE due to their low lattice conductivity (<2 $Wm^{-1}K^{-1}$)[10], high Seebeck coefficient[11], and the anticipation that its figures-of-merit be improved when thinned to TL scale, due to the increment of the density of states (DOS) when the band structure turns into MHVB.[5,6] Such thin PTMCs in theory would present zT as high as 2.0[5] and as good as the reference $Bi_2Te_3$ alloys. Nevertheless, the MHVB only appears in those materials in the form of very thin films, with less than 7 TLs[12] which results in difficulties to practical exploit its properties, including the growth or transfer of such thin films, the stability of the material in air[13,14], and the difficulties to measure tiny signals and properties at (Se-Ga(In)-Ga(In)-Se) tetralayer (TL) scale. For these reasons, despite the early theoretical interest, there are not experimental reports to date on the thermoelectric properties of few-layers PTMCs. Here we show that short-period GaSe/InSe superlattices (SL) share the same valence band properties. As a result, important electronic properties, in particular the thermoelectric figures-of-merit, that have already been calculated or measured in samples of few-layer thickness would be comparable in superlattices with the advantage of been amplified by the film thickness (number of SL periods) and facilitated application.

Starting from the indication that the MHVB is stronger in thinner films.[5,12] We performed calculations based on plane-waves density functional theory (DFT) to obtain the band structure of several InSe/GaSe SL SLs containing 1 to 3 TLs of GaSe in a 4 TL SL unit cell (Table 1). In this case, InSe replaces the vacuum of isolated layers as a barrier, due to type-II band alignments. In comparison to isolated layers, similar valley depth and peak k-point is obtained[5]. The SL with deepest valley (3 TL InSe / 1 TL GaSe SL) is then used in the following calculations.

Table 1. Comparison Γ-valley depth and peak k-point of the top of valence band for 4 TL SL unit cell heterostructures with ε-(P$\bar{6}$m2) stack using PBEsol functional.

|  | Γ-valley depth (meV) | Peak k-point (Å$^{-1}$) |
|---|---|---|
| 1 TL GaSe/3 TL InSe | 54 | 0.132 |
| 2 TL GaSe/2 TL InSe | 45 | 0.118 |
| 3 TL GaSe/1 TL InSe | 21 | 0.096 |

GaSe and InSe have several known polymorphs appearing depending on the growth method, the most common polytypes with non-centrossymmetric TL ($D_{3h}$) are named as ε-(P$\bar{6}$m2), β-(P$6_3$/mmc), γ-(R3m), and δ-(P$6_3$mc), based on different stacking sequences between the adjacent layers. [15-18] Nevertheless, we reported recently the first structural observation of a new GaSe and InSe polymorph (R$\bar{3}$m) (named as γ´-polymorph) characterized by a distinct atomic configuration with centrosymmetric TL ($D_{3d}$).[17,19,20] When with the same monolayer symmetry, polymorphs present small differences in band structure, electronic and optical properties particularities due to the presence or lack of special symmetries, however, usually these differences are not noticeable in most practical uses. Considering it, despite the γ´ and γ are the most common polytypes in our SL growth by MBE, we use the ε-stack and its centrosymmetric version P$6_3$/mc (named as ε´-polymorph) to compare between the $D_{3h}$ and $D_{3d}$ TLs SL, and these results should extend to the same TL symmetry polytypes. It is convenient not just because of the smaller unit cell (2 vs 3 TL) but also because the γ is rhombohedral with a different k-space, with A-A bandgap instead of Γ- Γ.[16] In the Figure 1 we show the comparison between $D_{3h}$ and $D_{3d}$ TLs SL band structure using Perdew-Burke-Ernzerhof (PBEsol)[21] generalized-gradient-approximation (GGA). We observe a small difference in the bandgap in the Γ- and M-point, a larger bandgap difference, and a spin degenerecence break in the K-point ($D_{3d}$ full Brillouin path in the supplementary information), however, more importantly, is the shape of the valence band. The centrosymmetric TL SL is the only one having MHVB. It suggests that the electronic hybridization between the TLs is reduced in the $D_{3d}$ polymorphs[22] and makes it a case closer to isolated TL.

Since the PBEsol functional is well-known to underestimate the bandgap, in the following calculations we used the modified Becke-Johnson exchange-correlation (mBJ) potential[23], which provides more accurate bandgaps and band structures in comparison to many-body theory (G$_0$W$_0$) and experimental data after the calibration of the *parameter c* that weights the Becke-Russel exchange[16]. In this way is not necessary to deal with scissor operators when the properties depend on the bandgap, as the properties of interest. For GaSe and InSe the *parameter c* was fixed (c = 0.838) after obtain the experimental band gap of β-InSe($D_{3h}$ TL) (1.25[15,16] eV). In the mBJ calculations (Figure 2a) the Γ-valley shows a depth of 17 meV and radius of 0.094 Å$^{-1}$. These are the main MHVB features, from which most of the characteristic material properties derive, and they are similar to the values found experimentally in isolated 2 TL GaSe[7]. The comparison between the functionals (Figure 2a), shows that these values can change slightly depending on the chosen functional, but the MHVB shape largely remains unaltered.

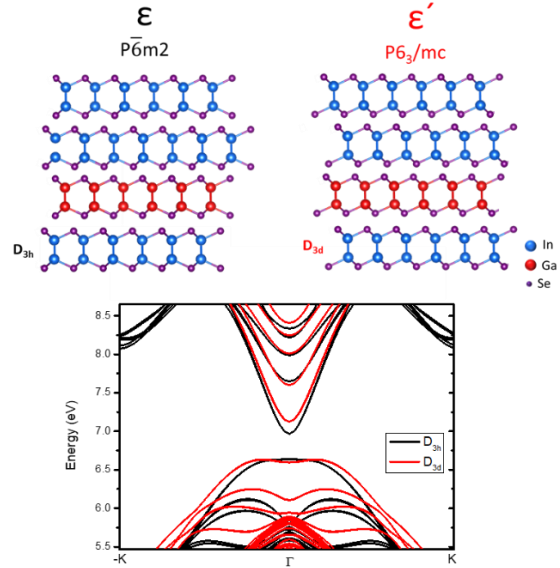

Figure 1. Comparison of crystal structure (top) of 1 TL GaSe/ 3 TL InSe with centrosymmetric ($D_{3d}$) and non-centrosymetric ($D_{3h}$) TLs in a ε-type stack along the [10$\bar{1}$0] plane, and its band structures around the Γ-point (bottom).

We proceed with our analysis taking InSe as a reference since this SL is actually ¾ InSe. Figure 2b makes this comparison with the calculated DOS. While the conduction band DOS is very similar on both materials, the SL exhibits van Hove singularities in the region where InSe has the standard parabolic band DOS due to MHVB. To state how it reflects in the TE properties, we performed a estimation of S, the power-factor ($S^2\sigma$), and zT (Figure 3) which is straightforward from the band structure and DOS, making use of Boltzmann equation[24]. The Seebeck coeficient, conductivity ($\sigma/\tau$) and electronic thermal conductivity ($k_e/\tau$) are calculated using BoltzTraP package and the output of DFT-mBJ. $\tau$ is the electron scattering time. The unknown values ($\tau$ and $\kappa_L$) can be obtained experimentally or from first principles calculations. In this work, $\tau$ (derivated from mean free path $\lambda_0$=25 nm) is extracted the literature[25] based on conductivity experimental data. The $\kappa_L$ (T) is obtained from first-principles calculations and is also in agreement with experimental values.[26] In fact, the zT and Seebeck coefficient obtained for bulk InSe are similar to experimental data[11,27] which validates these approximations.

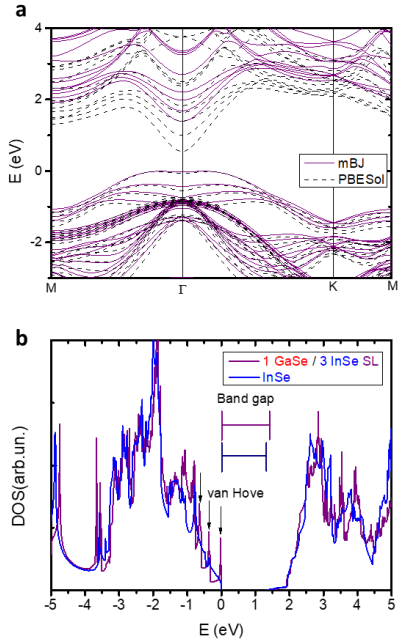

Figure 2. a) Calculated band structure of ε′-polymorph bulk 3 InSe TL / 1 GaSe TL superlattice (3/1 TL SL) using mBJ and PBESol functional b) Density of states (DOS) of ε′-polymorph bulk InSe and 1 TL GaSe/ 3 TL InSe SL.

Our calculations show that in comparison to bulk InSe the Seebeck coefficient barely changes for all temperatures and chemical potential considered. However, the singularities in the DOS caused by the MHVB result in a considerable increase in the electrical conductivity and the power factor without a corresponding increase in the $\kappa_e$. The outcome is that in-plane zT sharp increases (about 50% ) at room temperature in p-type SL. There is also a reduction of 200 K in the optimum temperature which makes it even more attractive for several applications at room-temperature. This increment in zT can be even bigger considering the improvement in $\kappa_L$ that occurs in SLs due to phonon confinement[28] in the heterostructure interfaces, and that was not considered in this first approximation.

We recently presented the epitaxial growth of InSe/GaSe heterostructures with atomically defined interfaces and $D_{3d}$ polymorphs using molecular beam epitaxy (MBE), which makes these superlattices readily available.[20] Few candidates for doping of PTMC already exist and can be used to achieve the chemical potential for peak zT.[29–32] While MBE would be a method appropriate for high-value wafers, other methods like chemical vapor deposition[33] (CVD), pulsed-laser deposition[34] (PLD) can be used to produce inexpensive large surfaces of this material in the future. Yet, The van der Waals (vdW) bond between the PTMC layers and the substrate facilitates the growth on a great variety of substrates and surface.

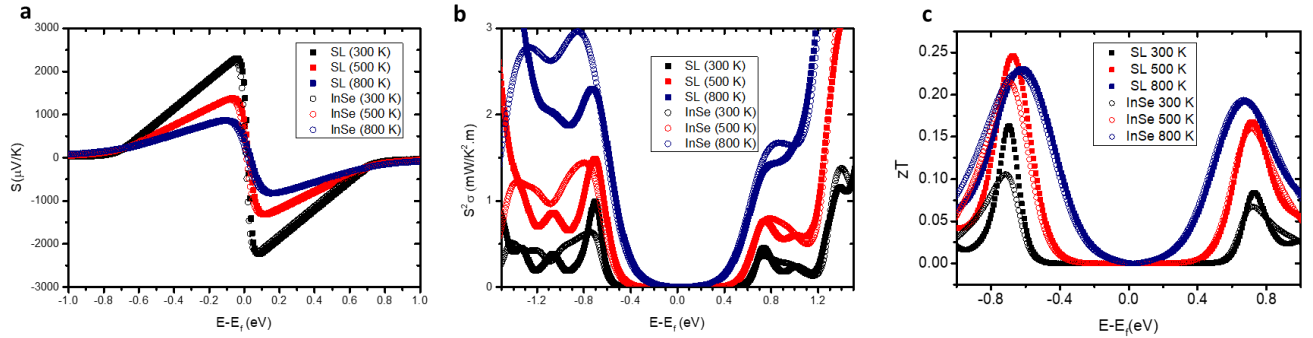

Figure 3. a)Seebeck coefficient, b) power factor c) zT thermoelectric figure-of-merit of InSe and 3/1 TL SL in function of temperature and chemical potential between 300 and 800 K.

In conclusion, using *ab initio* calculations we demonstrate that the in-plane zT can increases about 50% at room temperature in p-type on GaSe/InSe superlattices in comparison to bulk InSe due to the singularities in the DOS, which are present in the Mexican-hat valence band. MHVB exists only on non-centrosymmetric TL($D_{3d}$) SL, and a 1 TL GaSe/ 3 TL InSe SL presents a valance band shape which is similar to an isolated 2 TL GaSe. Similarly, this concept can be applied in other PTMCs, which are already considered as promising multifunctional materials, to enhance its performance as thermoelectric materials.

## Method

The *ab initio* calculations were done using density perturbation theory (DFT) within the Perdew-Burke-Ernzerhof (PBEsol)[21] generalized-gradient-approximation (GGA) and the modified Becke-Johnson exchange-correlation (mBJ meta-GGA) as implemented in Quantum Espresso[35,36] (QE) and LIBXC[37] using norm-conserving pseudopotentials[38]. We have optimized lattice parameters and atomic positions using PBEsol until forces in each atom are smaller than 0.01 eV/Å. For all systems, the basis-set cutoff energy is 140 Ry and the Brillouin zone integrated with 15x15x6 Γ-centered Monkhorst-Pack grid of k-points for bulk materials (β- ε- polytypes) and 15x15x3 grids for the 2 unit cells superlattices in the self-consistent calculations with convergence criteria of 1x10$^{-8}$ eV. We used twice the number of points (31x31x10 and 31x31x6) in non-self-consistent calculations for DOS and transport properties using BoltzTraP[24], which interpolate each k-point with 5 more intermediary points.

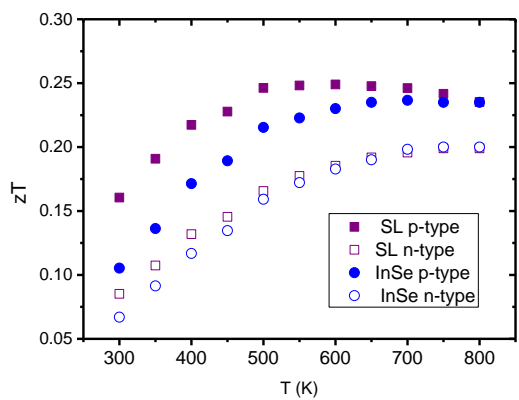

Figure 4. Comparison of Peak in-plane zT of n- and p-type bulk InSe and 1 TL GaSe/ 3 TL InSe SL between 300 K and 800 K.

## ACKNOWLEDGMENT

The computations were performed on the Tirant III cluster of the Servei d'Informàtica of the University of Valencia (project vlc82) thanks to Alejandro Molina-Sánchez and the Darwin cluster of INL-International Iberian Nanotechnology Laboratory thanks to Joaquín F. Rossier.

*Supplementary Information*

# Sharp enhancement on thermoelectric figure-of-merit of post-transition metal chalcogenides (PTMCs) using hetero-structures with Mexican-hat valence band

Marcel S. Claro[*,1,2]

[1] *INL-International Iberian Nanotechnology Laboratory, Av. Mestre José Veiga s/n, 4715-330 Braga, Portugal.*

[2] *Centro Singular de Investigación en Química Biolóxica e Materiais Moleculares (CiQUS), Departamento de Química-Física, Universidade de Santiago de Compostela, Santiago de Compostela 15782, Spain*

e-mail: marcel.santos@usc.es

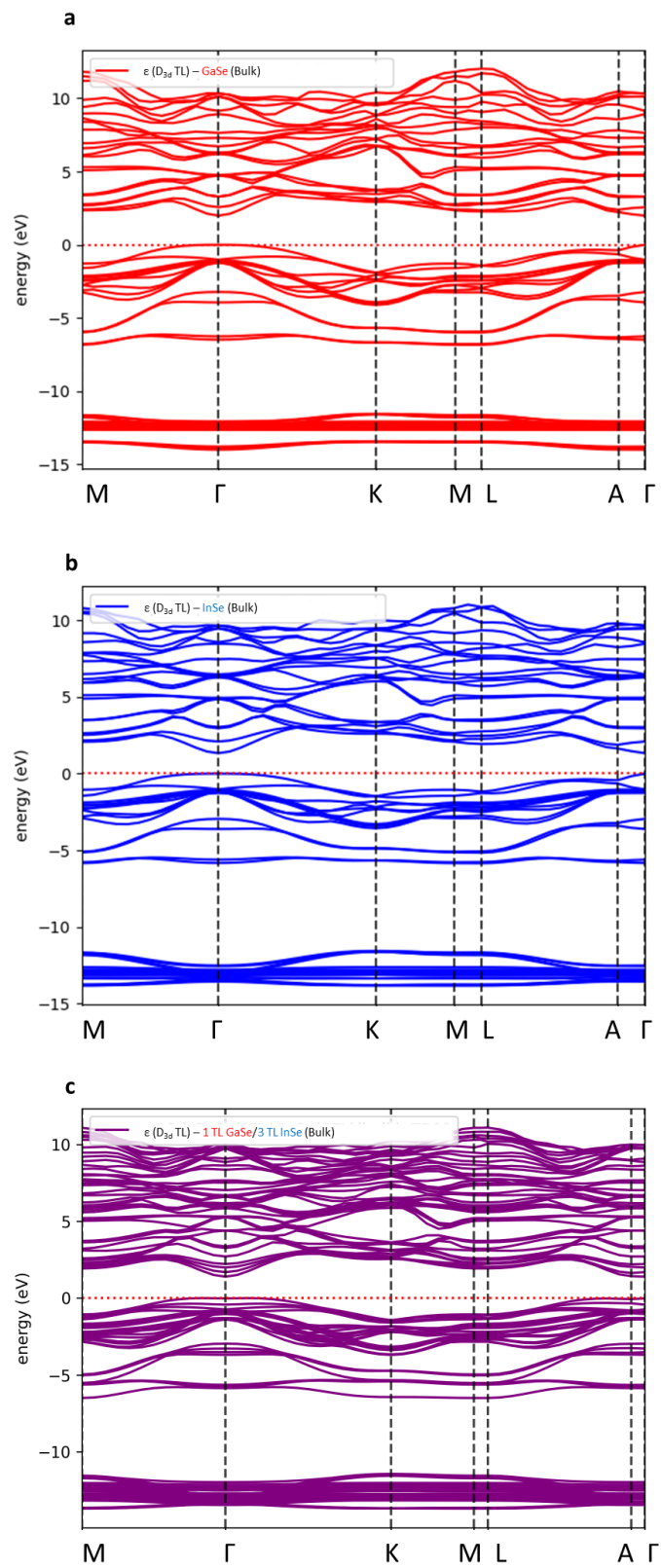

Figure S1. Band structure of P6$_3$/mc polytype of GaSe (a), InSe (b) and superlattice (c).